\documentclass{mn2e}
\usepackage{hyperref}
\usepackage{thumbpdf}
\usepackage{graphicx}
\usepackage{txfonts}
\begin{document}

\title[Radio and near-infrared observations of WKB 0314$+$57.8]{Radio and
near-infrared observations of the steep spectrum Galactic plane radio source
WKB 0314$+$57.8}

\author[D.~A.\ Green et al.]{D.~A.\ Green,$^1$\thanks{email: {\tt
   D.A.Green@mrao.cam.ac.uk}} M.\ Lacy,$^2$ S.\ Bhatnagar,$^{3,4}$
   E.~L.\ Gates,$^5$ and P.~J.\ Warner$^1$\\
$^1$ Mullard Radio Astronomy Observatory, Cavendish Laboratory,
     Madingley Road, Cambridge CB3 0HE\\
$^2$ Spitzer Science Center, MS220-6, California Institute of Technology,
     1200 East California Boulevard, Pasadena, CA 91106, USA\\
$^3$ National Centre for Radio Astrophysics, Pune 411 007, India\\
$^4$ National Radio Astronomy Observatory, 1003 Lopezville Road,
     Socorro, NM 87801, USA\\
$^5$ University of California Observatories/Lick Observatory,
     P.O.\ Box 85, Mount Hamilton, CA 95140, USA}

\maketitle

\begin{abstract}
Radio and near-infared observations towards the steep spectrum Galactic plane
radio source WKB 0314$+$57.8 are presented, in order to clarify the nature of
this source. The radio observations include archival and survey data, together
with new Giant Metrewave Radio Telescope observations at 617~MHz. The
near-infrared observations are in the $J$ and $K$ bands, from the Gemini
instrument on the Shane 3-m telescope. The radio observations show that WKB
0314$+$57.8 is extended, with an very steep spectrum (with ${\rm flux\ density}
\propto {\rm frequency}^{-2.5}$ between $\approx 40$~MHz and $\approx
1.5$~GHz). The colour--magnitude diagram constructed from near-infrared
observations of the field suggests the presence of a $z \approx 0.08$ galaxy
cluster behind the Galactic plane, reddened by about 6 magnitudes of visual
extinction. Although the steep spectrum source has no obvious identification,
two other radio sources in the field covered by the near-infrared observations
have tentative identifications with galaxies. These observations indicate that
WKB 0314$+$57.8 is a relic source in a cluster of galaxies, not a pulsar.
\end{abstract}


\section{Introduction}

The radio source WKB 0314$+$57.8 was first catalogued by Williams, Kenderdine
\& Baldwin (1966), with a flux density of 43~Jy at 38~MHz. This source is close
to the Galactic plane, near $l=141\fdg5$, $b=0\fdg5$, and also appears as a
relatively bright source in other low frequency radio surveys (e.g.\ the Clark
Lake surveys at 26.3~MHz and 30.9 MHz by Viner \& Erickson 1975 and Kassim 1988
respectively; the UTR-2 survey at several frequencies between 12.6 and 25~MHz
by Braude et al.\ 1988; the 8C survey at 38~MHz by Rees 1990 and Hales et al.\
1995). Williams et al.\ noted that although this source was not in the 4C
interferometric survey, it was in the 4CT `pencil beam' survey of Crowther \&
Clarke (1966). However, this source is {\em not} listed in Crowther \& Clarke,
nor in the published 4CT survey catalogue (Caswell \& Crowther 1969). It {\em
is} listed -- as `4C 57.06A' -- in Caswell (1966), with a flux density of 3~Jy
at 178~MHz, which implies a very steep radio spectral index, $\alpha$ of
greater than 1.5 (here $\alpha$ is defined in the sense that flux density, $S$,
scales with frequency as $\nu^{-\alpha}$). Caswell also noted that the flux
density of this source as observed in the 4C interferometric survey on a
469$\lambda$ baseline is 0.57 of the flux density detected with the 4CT pencil
beam observations, which implies the source is extended on the scale of several
arcminutes. Erickson \& Kassim (1985) noted that the steep spectrum of the
source, and its proximity to the Galactic plane, make it a pulsar candidate,
although the lack of scintillations (see, for example, Purvis et al.\ 1987)
argue against a pulsar identification, as does the fact noted above that the
source is several arcmin in extent. Subsequently this source has also been
observed in other radio surveys, and identified as an unusually steep spectrum
source. Green (1991) identified a source, 26P 218, observed 408~MHz in a
Galactic plane survey centred at $l=140^\circ$, $b=0^\circ$ as having a steep
radio spectrum, and Lacy (1992) identified the source as the steepest
low-frequency spectrum of any source in both the 8C (at 38~MHz) and 6C (at
151~MHz) catalogues. Both authors favoured identifying this source as an
unusually steep spectrum relic or halo source in a cluster, rather than a
pulsar, although the observations then available meant this was not a certain
conclusion. In particular, its position close to the Galactic plane means that
optical identification is difficult.

Here we present additional radio and infrared observations which, together with
observations from the literature or archives, confirm that this source is a
steep spectrum cluster source, not a pulsar. Details of the radio and infrared
observations are given in Sections~\ref{s:radio} and \ref{s:infrared}
respectively, are these are discussed in Section~\ref{s:discussion}.

\section{Radio Observations}\label{s:radio}

\subsection{GMRT Observations at 617~MHz}

The field of WKB 0314$+$57.8 was observed with the Giant Metrewave Radio
Telescope (GMRT; see Rao 2002) at frequencies near 618~MHz, on 2001 December 13
for about 2.3 hours. These observations were made during the commissioning of
the telescope, and typically about 25 of the 30 antennas of the GMRT were
available. The observations were made with a single 16~MHz sideband, centred at
618~MHz, using 128 spectral channels, for both right and left Stokes parameter.
The bright calibrator sources 3C48 and 3C286 were observed at the beginning and
end of the observing session respectively, to provide the flux density and
bandpass calibration. A secondary calibrator, 0432$+$416, was observed for
$\approx 4$ min every $\approx 30$ min to monitor the overall amplitude and
phase stability of the telescope. The data were calibrated using standard
procedures using classic AIPS (Bridle \& Greisen 1994). A few channels near the
centre of the band were inspected, and obvious interference was flagged. These
central channels were then averaged, applying an antenna-based bandpass
correction derived from the bright sources 3C48 and 3C286. The observations of
the secondary calibrators were then used to measure the antenna-based amplitude
and phase variations throughout the observations. The overall flux density
scale was set by the observations of the primary calibrators. The calibrated
data were inspected, and further interference was identified and flagged.
Several channels at each end of band were omitted, due to uncertain bandpass
corrections, and the resulting data -- with an effective bandwidth of 13.8~MHz
centred at 617.3~MHz -- were averaged into just 10 channels, to avoid bandwidth
smearing problems.

\begin{figure}
\centerline{\includegraphics[width=8.5cm]{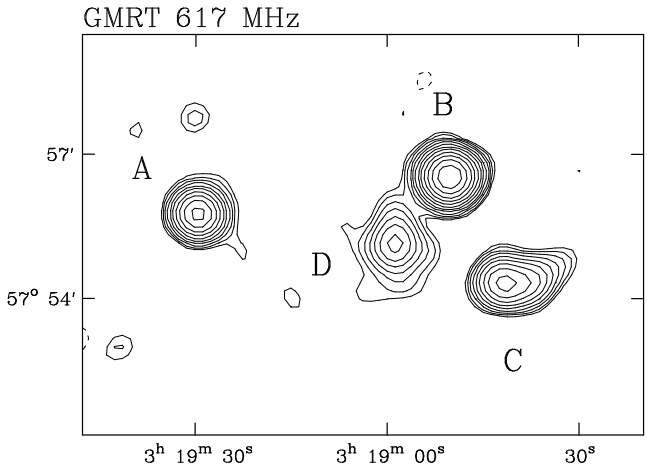}}
\caption{The field of WKB 0314$+$57.8 at 617~MHz, observed with the
GMRT, with a resolution of $41.4 \times 39.9$ arcsec$^2$ (at a position angle of
$18\fdg5$). Contours are at $2 \times \sqrt{2^n}$ mJy beam$^{-1}$ ($n=0, 1,
2, \dots$).\label{f:gmrt-low}}
\end{figure}


The GMRT consists of 30 antennas each 45~m in diameter, 12 in a central region
$\approx 1$~km in extent, with the others in three arms, providing baselines up
to $\approx 25$~km, i.e.\ $\approx 50$~k$\lambda$ at 617~MHz, with a primary
beam response of $\approx 0.7$ deg full width half maximum (FWHM). Fig.~\ref{f:gmrt-low} shows a
low-resolution ($\approx 40$ arcsec) image of WKB 0314$+$57.8 from the GMRT
observations -- which is comparable in resolution to other images shown below
-- by using only data from baselines shorter than 4 k$\lambda$. The image is
Stokes parameter I, from both the right and left circular polarisations
observed, and the noise is $\approx 0.5$ mJy beam$^{-1}$. The coordinates
are J2000.0, which are used for all images presented here.
Fig.~\ref{f:gmrt-low} shows three relatively bright sources around a fainter
extended source, which we designate A, B, C, and D respectively. (A higher
resolution, $\approx 10$ arcsec, image of these sources is also presented
below, in Fig.~\ref{f:ids}.)

\begin{figure}
\centerline{\includegraphics[width=8.5cm]{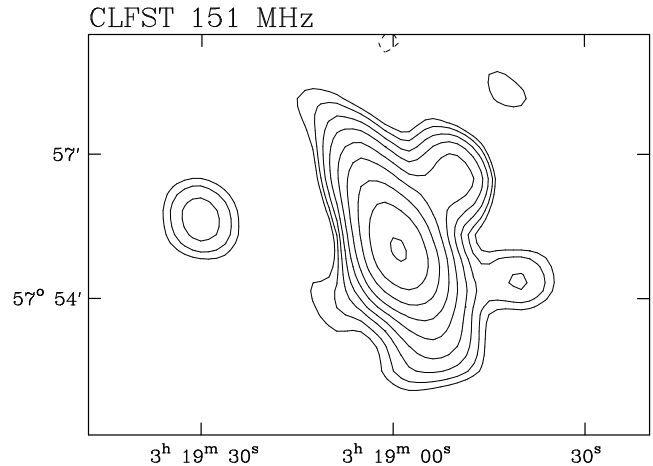}}
\caption{The region around WKB 0314$+$57.8 observed with the CLFST at 151~MHz.
The resolution is $83 \times 70$ arcsec$^2$ (NS $\times$ EW). Contours are at
$50 \times \sqrt{2^n}$ mJy beam$^{-1}$ ($n=0, 1, 2, \dots$).\label{f:clfst}}
\end{figure}

\subsection{CLFST observations at 151~MHz}

Lacy (1992) presented an image of WKB 0314$+$57.8 made with the Cambridge
Low-Frequency Synthesis Telescope (CLFST, see McGilchrist et al.\ 1990; Rees
1990) at 151~MHz which we reproduce here. A total of eight 12-hour CLFST
observations of the source were made, between 1989 September 2 and 1990 October
13. These observations were processed using standard procedures used for CLFST
analysis (e.g.\ McGilchrist et al.), including correction for time varying
ionospheric phase gradients across the telescope. The final image,
Fig.~\ref{f:clfst} -- which has a noise of $\approx 18$ mJy beam$^{-1}$ -- is a
weighted sum of the eight observations. This image is dominated by source D,
which is clearly extended.

\begin{figure}
\centerline{\includegraphics[width=8.5cm]{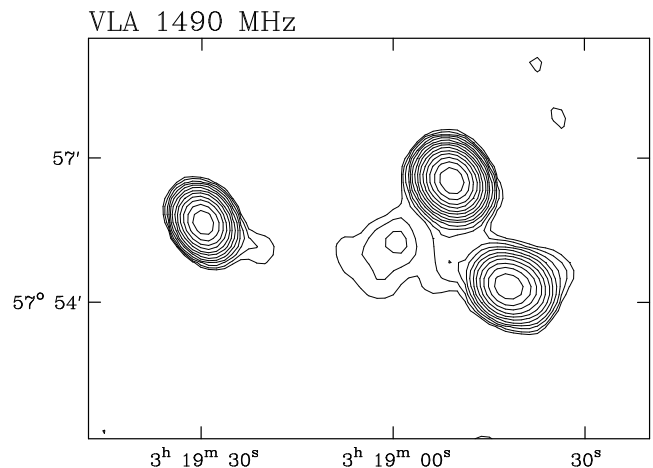}}
\caption{VLA D-array observations of WKB 0314$+$57.8 at 1490~MHz,
with a resolution of $48.7 \times 37.0$ arcsec$^2$ (at a position angle of
$26\fdg9$). Contours are at $0.5 \times \sqrt{2^n}$ mJy beam$^{-1}$ ($n=0, 1,
2, \dots$).\label{f:vla}}
\end{figure}

\begin{figure}
\centerline{\includegraphics[width=8.5cm]{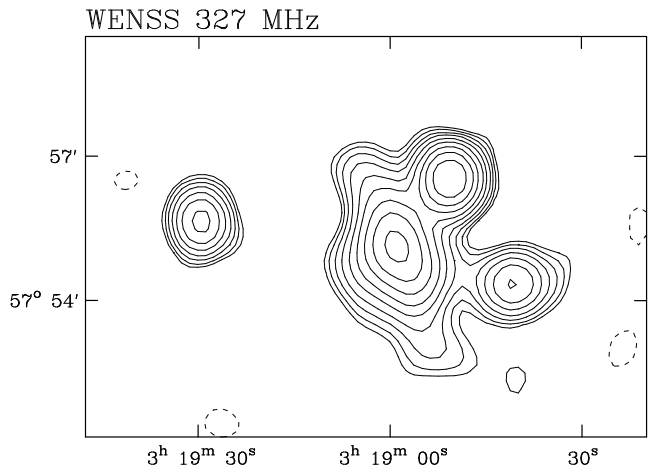}}
\caption{The region around WKB 0314$+$57.8 from the WENSS survey at 327~MHz.
The resolution is $64 \times 54$ arcsec$^2$ (NS $\times$ EW). Contours are at
$-10, 10 \times \sqrt{2^n}$ mJy beam$^{-1}$ ($n=0, 1, 2, \dots$; negative
contours are dashed).\label{f:wenss}}
\end{figure}

\subsection{Archive and Survey Radio Observations}

In addition to the GMRT and CLFST radio observations of WKB 0314$+$57.8
presented above, there are several other sources of useful radio observations
available in archives, or recent surveys. WKB 0314$+$57.8 was observed with the
Very Large Array (VLA) of the National Radio Astronomy Observatory\footnote{The
National Radio Astronomy Observatory is a facility of the National Science
Foundation operated under cooperative agreement by Associated Universities,
Inc.} in D-array at 1.5~GHz for about 1.4 hours on 1990 January 25. We obtained
these observations from the online VLA archive, and processed them using
classic AIPS procedures. Fig.~\ref{f:vla} show the resulting synthesised image,
which has a r.m.s.\ noise of $\approx 0.05$ mJy beam$^{-1}$. This image was
made from data observed in two 50~MHz bands, centred at 1465 and 1515 MHz, so
is at an effective frequency of 1490~MHz. Sources A, B and C are relatively
bright and well defined in this image, but source D is faint, and extended, and
not well defined.

WKB 0314$+$57.8 is also included in the WENSS survey (Rengelink et al.\ 1997)
at 327~MHz. Fig.~\ref{f:wenss} shows the image of WKB 0314$+$57.8 from this
survey. In this image source D is relatively much brighter than at the higher
frequencies, and it along with sources A, B and C, is well defined.

\section{Near-infrared observations}\label{s:infrared}

The field of WKB 0314$+$57.8 was observed in the near-infrared $J$ and
$K$-bands with the Gemini instrument at Lick Observatory on 2002 August 27.
Gemini is a near-infrared imaging camera with long and short wavelength
channels operated at the Cassegrain focus of the Shane 3-m telescope. Radiation
entering the instrument is passed through a dichroic beam splitter into two
arms, short wavelengths ($< 2$ $\mu$m) are transmitted to a $256 \times 256$
NICMOS3 HgCdTe array from Rockwell International, and longer wavelengths to a
InSb array from SBRC, also with $256\times 256$ pixels. Both arrays have scales
of 0.7 arcsec pixel$^{-1}$, giving a $3\times 3$ arcmin$^2$ field of view. We
observed with a $J$-filter in the short arm and a $K$ filter in the long arm.
The two arms are not quite confocal, so we set the telescope focus to give the
best focus in $K$-band. We observed two pointings, one covering both sources C
and D, and one covering source B. We observed both pointings in a 9-point grid
for two minutes per point, using 24 coadds of 5 s exposures in $K$-band and 3
coadds of 40 s in $J$. The data were flux calibrated using the standards P138-C
and S754-C from the list of Persson et al.\ (1998).

The data were dark subtracted, flat-fielded using twilight flats. The {\sc
dimsum} package in IRAF (Stanford, Eisenhardt \& Dickinson 1995) was then used
to subtract a running median background image made from the nine frames closest
in time to each image, and to coadd the data into the final mosaics.
Astrometric plate solutions were derived using the positions of Two-Micron All
Sky Survey (2MASS) stars in the images, with the resulting astrometry accurate
to $\approx 0.2$ arcsec.

%
%
\begin{table}
\caption{Integrated flux densities of sources A, B, C and D (see
Fig.~\ref{f:gmrt-low} for the source positions).\label{t:abcd-fluxes}}
\begin{tabular}{ccccc}
source & \multicolumn{3}{c}{flux density/mJy}    \\
       &  WENSS  &  GMRT   &   VLA     \\
       & 327 MHz & 617 MHz & 1490 MHz  \\ \hline
  A    &    89   &   61    &    33     \\
  B    &   221   &  150    &    76     \\
  C    &   103   &   62    &    34     \\
  D    &   458   &   52    &     6     \\ \hline
\end{tabular}
\end{table}
%
%
%
%

\section{Discussion and Conclusions}\label{s:discussion}

The integrated flux densities of the sources A, B, C, and D from the WENSS
327~MHz, GMRT 617~MHz and VLA 1490~MHz observations are given in
Table~\ref{t:abcd-fluxes}. The WENSS flux densities are taken from the WENSS
catalogue, which lists source B, C, and D as components of a multi-component
source. The other flux densities are estimated from the images shown above.
Generally these flux densities are to be accurate to better than 10 per cent.
However, since source D is faint at 1490~MHz, and is extended (see
Fig.~\ref{f:vla}), its flux density is difficult to measure accurately. From
these flux densities, it is, however, clear that it is source D that has an
unusually steep spectrum, and it must be associated with the relatively bright
source observed in various surveys at lower frequencies.

\begin{table*}
\caption{Radio flux densities of WKB 0314$+$57.8 (see Fig.~\ref{f:spectrum} for
a plot of this spectrum).\label{t:fluxes}}
\begin{tabular}{rccll}
frequency & flux density & estimate error & source     & reference \\
 /MHz     &    /Jy       &      /Jy       &            &           \\ \hline
    16.7  &    83        &   10.0         & UTR-2      & Braude et al.\ (1988)                  \\
    26.3  &    88        &   10           & Clark Lake & Viner \& Erikson (1975)                \\
    30.9  &    53.1      &   10.6         & Clark Lake & Kassim (1988)                          \\
    38.0  &    58.6      &    6           & 8C         & Rees (1990); Hales et  al.\ (1995)     \\
    38.0  &    43        &    4.3         & WKB        & Williams, Kenderdine \& Baldwin (1966) \\
   151    &     4.38     &    0.5         & 6C         & Hales et al.\ (1993)                   \\
   178    &     3.0      &    0.5         & 4CT        & Caswell (1966)                         \\
   327    &     0.458    &    0.046       & WENSS      & Rengelink et al.\ (1997)               \\
   617    &     0.052    &    0.005       & GMRT       & this work                              \\
  1490    &     0.006    &    0.0012      & VLA        & this work                              \\ \hline
\end{tabular}
\end{table*}

\begin{figure}
\centerline{\includegraphics[width=8.5cm]{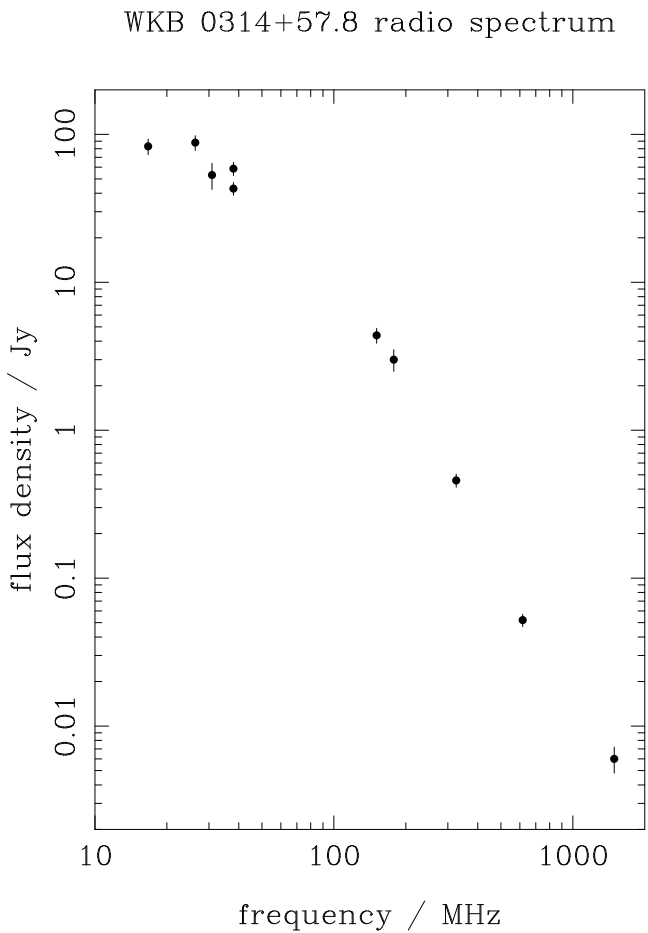}}
\caption{Radio spectrum of WKB 0314$+$57.8 (see
Table~\ref{t:fluxes}).\label{f:spectrum}}
\end{figure}

\begin{figure}
\centerline{\includegraphics[width=8.5cm]{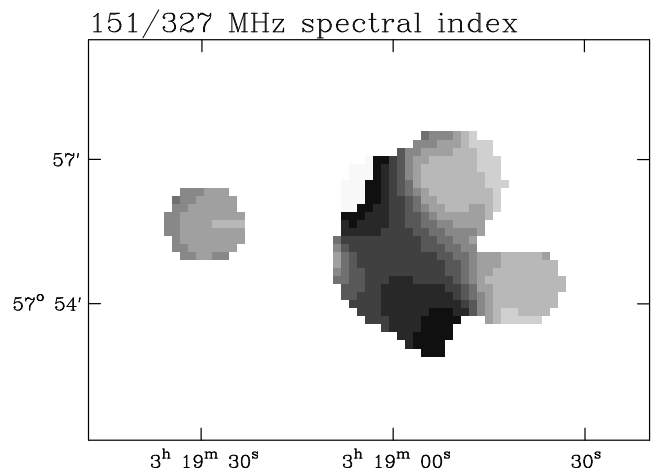}}
\caption{Radio spectral index of sources A, B, C and D, between 151-MHz (the
CLFST observations shown in Fig.~\ref{f:clfst}) and 327-MHz (the WENSS
observations shown in Fig.~\ref{f:wenss}, convolved to the resolution of the
CLFST observations). The greyscale transitions are, light to dark, at $0.7,
1.0$\dots$ 3.1$ (i.e.\ sources B, C have a spectral index that is between 0.7
and 1.0.)\label{f:alpha}}
\end{figure}

The radio spectrum of WKB 0314$+$57.8, now associated with source D, is shown
in Fig.~\ref{f:spectrum}, using the flux densities given in
Table~\ref{t:fluxes}. For frequencies of 327~MHz and above the flux densities
used are for source D only, whereas for the lower frequency observations --
with poor resolution -- the catalogued sources will be included contributions
from the other nearby sources.\footnote{For example, the steep spectrum source,
26P 218, in this region that was noted by Green (1991), was identified from
observations made at 408~MHz with the Dominion Radio Astrophysical
Observatory (DRAO) Synthesis Telescope with a
resolution of $3.4 \times 3.8$ arcmin$^2$ (EW $\times$ NS). Given this
resolution of these observations, and the proximity of other sources, the
408-MHz flux density of 26P 218 does not correspond solely to source D, and
hence it is not included in Fig.~\ref{f:spectrum} and Table~\ref{t:fluxes}.}
However, given the very steep spectrum of source D, these contributions are not
large at the lower frequencies. The flux densities are as published, and no
attempt has been made to place them on a consistent flux density scale, which
is not straightforward at the lower frequencies. Nevertheless, even allowing
for uncertainties in the flux density scales of the observations,
Fig.~\ref{f:spectrum} clearly shows that WKB 0314$+$57.8 has an remarkably
steep radio spectrum from $\sim 40$~MHz to $\sim 1.5$~GHz, with $\alpha \approx
2.5$. At lower frequencies, the spectrum of this source turns over, which is to
be expected from Galactic absorption.

Given the unusually steep radio spectrum of this source, the obvious
interpretation as to its nature is either as a pulsar, or as a cluster source,
as both of these types of object are known to have very steep radio spectra. In
the case of pulsars the emission mechanism produces an intrinsically steep
spectrum (e.g.\ Sieber 1972; Maron et al.\ 2000), whereas for cluster radio
sources, the synchrotron emission from electron populations confinded by the
intracluster medium produces steeper spectra with age (e.g.\ Slingo 1974;
Cordey 1987; Slee et al.\ 2001 respectively). Despite being close to the
Galactic plane -- which suggests a pulsar identification -- the fact that WKB
0314$+$57.8 is clearly extended means that it cannot be a pulsar, but is
instead likely to be a cluster source, that by chance is close to the Galactic
plane. Fig.~\ref{f:alpha} shows a spectral index image of sources A, B, C and
D, from the CLFST and WENSS observations. This shows that the spectral index of
source D gradually steepens to the north and south, away from its brightest
emission. This variation, and the structure of source D, suggests it is a FR
class I source (Fanaroff \& Riley 1974), in which spectral ageing of the
particles in the lobes produces the very steep radio spectrum seen.

\begin{figure}
\centerline{\includegraphics[width=8.5cm]{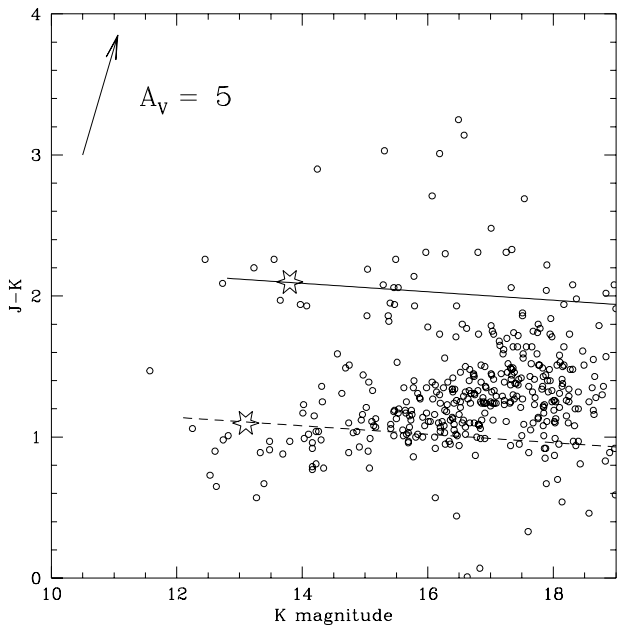}}
\caption{The near-IR colour--magnitude diagram for the field of WKB
0314$+$57.8. The dashed line indicates the colour--magnitude relation for a
$z=0.08$ unreddened cluster, the solid line that for one reddened by
$E(J-K)=1.0$ ($A_V=5.9$). The stars indicate the position of a galaxy with
magnitude $M^{*}$ on each of the colour--magnitude relations. The reddening
vector for $A_V=5$ is shown at top left.\label{f:mir}}
\end{figure}

Our infrared observations in this direction are less affected by absorption
than optical observations would be, allowing us to make a deep search of any
host cluster of  WKB 0314$+$57.8. The reduced mid-infrared mosaics were run
through SExtractor (Bertin \& Arnouts 1996) to produce object catalogues, and
sources in the $J$ and $K$-band images were matched on the basis of position.
Due to the poor seeing we have made no attempt at star-galaxy separation.
Figure~\ref{f:mir} shows the colour--magnitude diagram for the objects in the
field. Most objects have colours consistent with them being moderately-reddened
stars, but there is evidence for a cluster colour--magnitude sequence with
$J-K\approx 2$. This is, of course, much redder than seen in low redshift
clusters where $J-K\approx 1$, but reddening through the Galactic plane is to
be expected. We have made an estimate of the redshift of the cluster by
assuming that the brightest galaxy on the colour--magnitude sequence is the
Brightest Cluster Galaxy (BCG). We assume an intrinsic (unreddened) colour of
$J-K\approx 1.1$ (for a $z=0.05{-}0.10$ BCG) compared to a measured $J-K \approx
2.1$ and thus estimate the colour excess in $J-K$, $E(J-K) \approx 1.0$. Using
the Milky Way extinction law of Cardelli, Clayton \& Mathis (1989) we estimate
the extinction in $K$-band, $A_K$, to be 0.7 magnitudes (corresponding to a
visual extinction, $A_V$, of 5.9), which would make the intrinsic BCG magnitude
$K=12.0$. We then use the $K{-}z$ relation of Brough et al.\ (2002) to estimate
a redshift of $0.08\pm 0.02$ for the BCG.

\begin{figure}
\centerline{\includegraphics[angle=270,width=7cm,clip=]{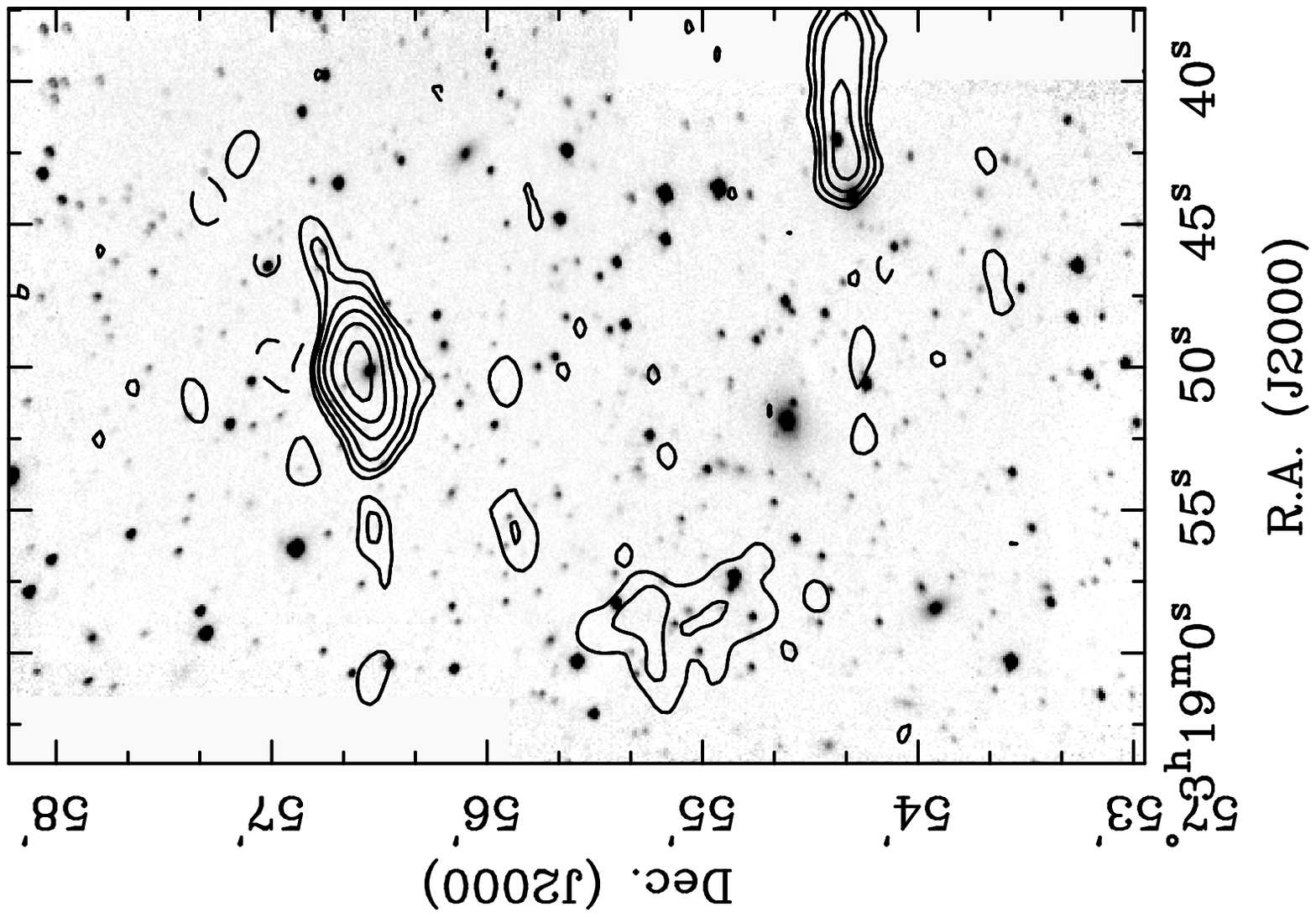}}
\caption{Contours of radio emission at 617~MHz, from the GMRT observations,
overlaid on a $K$-band image. The contours are $\pm 2^n$ mJy beam$^{-1}$ ($n=0,
1, 2\dots$), with negative contours dashed.\label{f:ids}}
\end{figure}

On Fig.~\ref{f:mir} we also plot the $z=0.08$ colour--magnitude relation using
the mean of the slopes measured by Bower, Lucey \& Ellis (1992) for the Coma
and Virgo clusters, $k$-corrected using elliptical galaxy corrections from the
{\sc pegase} model (Fioc \& Rocca-Volmerange 1997). We plot both the unreddened
$z=0.08$ relation (plotted from a typical brightest cluster galaxy one
magnitude brighter than $M^{*}$ and fainter) and one reddened by $E(J-K)=1.0$,
$A_K=0.7$ which appears to match the observed sequence fairly well.

The low resolution of the available radio data, and the crowded nature of the
field in the near-infrared make identification of the radio sources uncertain.
Figure~\ref{f:ids} shows an overlay of contours of a higher resolution 617-MHz
radio image on a greyscale of the $K$-band image. The radio image has a
resolution of $9.8 \times 8.1$ arcsec$^2$ (at a position angle of $98\fdg7$), and a
noise of $\approx 0.2$ mJy beam$^{-1}$. At this higher resolution, sources B
and C are clearly resolved, and are each extended in approximately the
east--west direction. Source B is close to a $K=13.7$ galaxy with $J-K=1.9$ at
$03^{\rm h} 18^{\rm m} 50\fs1$, $+57^{\circ} 56' 31''$. The identification of
source C is uncertain, there is a $K=13.2$, $J-K=2.2$ galaxy at the eastern
extreme of the radio contours at $03^{\rm h} 18^{\rm m} 44\fs1$, $+57^{\circ}
54' 19''$ which may be at the head of a head-tail source. A bright ($K=13.5$)
stellar object with $J-K=0.9$ is located within the radio contours at $03^{\rm
h} 18^{\rm m} 42\fs1$ $+57^{\circ} 54' 23''$, but this is most likely a
foreground star. Source D is not clearly identified with any particular cluster
galaxy.

Assuming WKB 0314$+$57.8 is associated with a cluster at $z\approx 0.08$, then
at 178~MHz its luminosity would be $\approx 8 \times 10^{25}$ W Hz$^{-1}$ (for
a Hubble constant of 50 km s$^{-1}$ Mpc$^{-1}$, as used by Fanaroff \& Riley
1974 and Zirbel \& Baum 1995 below). This is somewhat high compared with
most FR class I sources (e.g.\ Fanaroff \& Riley), but due to its very steep
spectrum, at higher frequencies its luminosity is much lower, and is consistent
with other FR class I sources (e.g.\ Zirbel \& Baum). Also at this redshift its
linear size -- from the largest angular extent of $\approx 5$ arcmin at 151~MHz
(Fig.~\ref{f:clfst}) -- would be $\approx 700$~kpc, which is comparable with
with the physical sizes of other cluster sources (e.g.\ Slingo 1974).


In the ROSAT all-sky bright source catalogue (Voges et al.\ 1999) an X-ray
source (1RXS J031902.8$+$575504) is detected nearly coincident with source D.
This X-ray source is at $03^{\rm h} 19^{\rm m} 02\fs8$, $+57^\circ 55' 04''$,
with a position uncertainty of 16 arcsec, and has a count rate of 0.015
s$^{-1}$. Although unresolved by the 30-arcsec ROSAT beam, it is plausible that
this emission is from the galaxy cluster, heavily attenuated by the gas column
of the Milky Way. Using the PIMMS tool\footnote{{\tt
http://heasarc.gsfc.nasa.gov/Tools/w3pimms.html}} we estimated the X-ray flux
from the source in the $0.5{-}2$ keV band. We assume a thermal bremsstrahlung
spectrum and an absorption column of $10^{22}$ cm$^{-2}$, estimated from the
optical reddening and a standard Milky Way gas to dust ratio of $N_{\rm
H{\scriptscriptstyle\rm I}} = 6\times 10^{21} E(B-V)$, where $E(B-V)\approx
A_V/3.1$. These assumptions yield an X-ray flux estimate of $2\times 10^{-13}$
erg s$^{-1}$ cm$^{-2}$. (The flux in the absence of absorption is estimated to
be $9\times 10^{-13}$ erg s$^{-1}$ cm$^{-2}$.) At $z=0.08$, this corresponds to
an X-ray luminosity of $1.4 \times 10^{43}$ erg s$^{-1}$, equivalent to a
moderately-luminous galaxy cluster.

In conclusion, the extended nature of WKB 0314$+$57.8 at radio wavelengths, our
mid-infrared identification of a likely cluster in this direction, and the
ROSAT X-ray source lead us to conclude that this ultra steep radio source is a
cluster source, not a pulsar.

\section*{Acknowledgements}

We are grateful to Guy Pooley and David Titterington for help accessing the
archived CLFST observations. The GMRT is run by the National Centre for Radio
Astrophysics of the Tata Institute of Fundamental Research, India. This
research has made use of the VizieR catalogue access tool and he SIMBAD
database, operated at CDS, Strasbourg, France, and NASA's Astrophysics Data
System Bibliographic Services.


\end{document}